\documentclass[12pt]{article}
\usepackage{a4wide,epsfig,psfrag,amsmath,amssymb,cite,scalefnt}
\usepackage[dvipsnames]{xcolor}
\usepackage{amsmath,comment,braket}
\usepackage{hyperref}

\graphicspath{ {figs/} }

\newcommand{\F}[1]{H^{(c)\,{#1}}}
\newcommand{\FS}[1]{\widetilde{H}_S^{(c)\,{#1}}}
\newcommand{\PP}[1]{H^{({c}p)\,{#1}}}
\newcommand{\PS}[1]{\widetilde{H}_S^{({c}p)\,{#1}}}
\newcommand{\PPP}[1]{H^{(p)\,{#1}}}
\newcommand{\PPS}[1]{\widetilde{H}_S^{(p)\,{#1}}}

\newcommand{\lt}{\left}
\newcommand{\rt}{\right}
\newcommand{\no}{\nonumber}
\newcommand{\nn}{\nonumber\\}
\newcommand{\ov}{\overline}
\newcommand{\eq}[1]{Eq.~(\ref{#1})}
\newcommand{\eqsand}[2]{Eqs.~(\ref{#1}) and (\ref{#2})}

\newcommand{\Bbar}{\bar{B}}
\newcommand{\bbd}{\ensuremath{B_d\!-\!\Bbar{}_d\,}}
\newcommand{\bbs}{\ensuremath{B_s\!-\!\Bbar{}_s\,}}
\newcommand{\bbq}{\ensuremath{B_q\!-\!\Bbar{}_q\,}}
\newcommand{\bbms}{\bbs\ mixing}
\newcommand{\bbmd}{\bbd\ mixing}
\newcommand{\bbmq}{\bbq\ mixing}

\newcommand{\lqcd}{\Lambda_{\rm QCD}} 
\newcommand{\dm}{\ensuremath{\Delta M}}
\newcommand{\dg}{\ensuremath{\Delta \Gamma}}
\newcommand{\logOne} {L_1} % means log(mu1^2/mb^2)
\newcommand{\logTwo} {L_2} % means log(mu2^2/mb^2)

\allowdisplaybreaks

\begin{document}

\title{\vskip-3cm{\baselineskip14pt
    \begin{flushleft}
      \normalsize P3H-21-040, TTP21-015
    \end{flushleft}} \vskip1.5cm 
  Two-loop QCD penguin contribution to the width difference
  in $B_s-\bar{B}_s$ mixing
}
\author{
  Marvin Gerlach, Ulrich Nierste, Vladyslav Shtabovenko,
  \\
  and Matthias Steinhauser
  \\[1em]
  {\small\it Institut f{\"u}r Theoretische Teilchenphysik,
    Karlsruhe Institute of Technology (KIT)}\\
  {\small\it 76128 Karlsruhe, Germany}  
}
  
\date{}

\maketitle

\thispagestyle{empty}

\begin{abstract}
  We consider two-loop QCD corrections to the element $\Gamma_{12}^q$ of the
  decay matrix in  $B_q-\bar{B}_q$ mixing, $q=d,s$, in the leading power
  of the Heavy Quark Expansion. The calculated
  contributions  involve one current-current and one penguin operator
  and constitute the next step towards a theory prediction for 
  the width difference $\Delta\Gamma_s$ matching the precise
  experimental data.  We present compact analytic results for all
  matching coefficients in an expansion  in $m_c/m_b$ up to second
  order. Our new corrections are comparable in size to the current
  experimental error and slightly increase  $\Delta\Gamma_s$.
\end{abstract}

%- }}}

\thispagestyle{empty}

\newpage

%- {{{ Introduction and notation:

\section{Introduction}
In particle collisions $B_q$ mesons, where $q=d,s$ labels the flavour of the
light valence quark, are produced as flavour eigenstates. This means that they
are either meson or antimeson, with beauty quantum number $B=1$ or $B=-1$,
respectively.  Subsequently, this pure $B_q$ or $\bar B_q$ state evolves into
a quantum-mechanical superposition of $B_q$ and $\bar B_q$ following the time
evolution of damped oscillations. Two accidental features of the Standard
Model (SM) permit the precise study of \bbq\ oscillation in modern
experiments: First, the smallness of the element $V_{cb}$ of the
Cabibbo-Kobayashi-Maskawa (CKM) matrix implies a large $B_q$ lifetime of around
1.5\,ps, which makes decay-time dependences experimentally
observable. Second, the heaviness of the top quark enhances the \bbmq\ box
diagram, which governs the \bbmq\ amplitude, to a level that the oscillation
frequency is in the same ballpark as the $B_q$ lifetime.

\bbmq{} is described by the $2\times2$ matrix $M^q - i \Gamma^q/2$
with the hermitian mass and decay matrices $M^q$ and $\Gamma^q$,
respectively.  Diagonalizing $M^q - i \Gamma^q/2$ leads to a ``heavy''
(H) and a ``light'' (L) mass eigenstate which are commonly
denoted by $B_H^q$ and $B_L^q$, respectively, and have masses
  $M_{H,L}$ and widths $\Gamma_{H,L}$.  The oscillation phenomena
involve the three quantities $|M_{12}^q|$, $|\Gamma_{12}^q|$ and
$\mbox{arg}(-M_{12}^q/\Gamma_{12}^q)$ which are related to the
experimentally accessible quantities
\begin{eqnarray}
  \Delta M_q &=& M_H^q - M_L^q\,,\nonumber\\
  \Delta \Gamma_q &=& \Gamma_L^q - \Gamma_H^q\,,\nonumber\\
  a_{\rm fs}^q &=& \mbox{Im} \frac{\Gamma_{12}^q}{M_{12}^q}\,, \label{eq:dgdmafs}
\end{eqnarray}
where the CP asymmetry in flavour-specific decays, $a_{\rm fs}^q$,
is typically measured in semileptonic decays.
$\dm_q$ and $\dg_q$ are related to the elements of the mass and
decay matrices as
\begin{eqnarray}
  \dm_q & \simeq & 2 |M_{12}^q|, \qquad\qquad
     \frac{\Delta\Gamma_q}{\Delta M_q} \;=\; - \mbox{Re}\frac{\Gamma_{12}^q}{M_{12}^q}\,.             
  \label{eq:dgdm}
\end{eqnarray}
In the Standard Model (SM) the phase between $-\Gamma_{12}^q$ and
$M_{12}^q$ is small, so that $ a_{\rm fs}^q$ is much smaller than
$\Delta\Gamma_q/\Delta M_q$ and $\dg_q \simeq 2|\Gamma_{12}^q|$.

$M_{12}^q$ is a $\Delta B=2$ amplitude probing virtual effects of physics
beyond the SM (BSM physics) up to mass scales of several 100 TeV. By contrast,
$\Gamma_{12}^q$ is sensitive to new physics in $\Delta B=1$ transitions. While
$\Gamma_{12}^q$ probes much lower scales than $M_{12}^q$, it is instead
sensitive to effects of feebly coupled BSM particles which are light enough to
be produced in $B_q$ decays. Such particles are predicted in theories
addressing the strong CP problem \cite{Calibbi:2016hwq,MartinCamalich:2020dfe}
or as members of the dark sector, see e.g.\ Ref.~\cite{Elor:2018twp} for a
baryogenesis mechanism utilising \bbmq\ and $B_q$ decays into dark matter.

In this paper we calculate QCD corrections to 
$\Gamma_{12}^q$ in the SM
needed to better predict both $\Delta\Gamma_q/\Delta M_q$ and
$a_{\rm fs}^q$. Currently, better theory predictions are needed in
the case of \bbms\ to be competitive with the precise experimental values
\begin{eqnarray}
  \dm^{\rm exp}_s &=& (17.757 \pm 0.007_{\rm (stat)} \pm 0.008_{\rm (syst)})
            \; \mbox{ps$^{-1}\qquad$\cite{Aaij:2020cax}} \nn
  \dg^{\rm exp}_s  &=&  (0.085 \pm 0.004)\;
                      \mbox{ps}^{-1}. \qquad
                      \mbox{\cite{hfag}}  \label{eq:exp}
\end{eqnarray}  
Furthermore, there is steady progress with measurements of $\dg_s$ at
LHCb \cite{Aaij:2019vot}, CMS \cite{Sirunyan:2020vke}, and ATLAS
\cite{Aad:2020jfw}.

For the calculation of $\Gamma_{12}^q$ one employs a special operator product
expansion, the Heavy Quark Expansion (HQE), which treats the $b$ quark mass
$m_b$ as a hard scale. In this way one expresses $\Gamma_{12}^q$ as a
simultaneous expansion in $\lqcd/m_b$ and $\alpha_s(m_b)$. Each term of the
$1/m_b$ expansion involves perturbative coefficients multiplying hadronic
matrix elements of local $\Delta B=2$ operators.  Next-to-leading logarithmic
order (NLO) QCD corrections at leading power in
$1/m_b$ have been computed in
Refs.~\cite{Beneke:1998sy,Ciuchini:2003ww,Beneke:2003az,Lenz:2006hd}.  The
$1/m_b$ contribution is known to leading order in $\alpha_s$
\cite{Beneke:1996gn}. The uncertainty resulting from the truncation of the
perturbative series of the currently known SM prediction for $\dg_s$ is larger
than the experimental error in \eq{eq:exp}, which calls for the calculation of
higher-order QCD contributions.

First steps towards next-to-next-to-leading order (NNLO) have been undertaken
in Ref.~\cite{Asatrian:2017qaz} where the fermionic corrections of order
$\alpha_s^2 N_f$, where $N_f=5$ is the number of active quark flavours, have
been computed including linear terms in the expansion in $m_c/m_b$.  Note that
this calculation cannot be used to obtain $a_{\rm fs}^q$, which is
proportional to $m_c^2/m_b^2$.  In this paper we denote any
${\cal O}(\alpha_s)$ contribution to $\Gamma_{12}^q$ as ``NLO'', irrespective
of the Wilson coefficients involved. This complies with the commonly used
notation in connection with higher-order QCD calculations, but differs from
the language used in previous papers on $\Gamma_{12}^q$, in which the small
$\Delta B=1$ penguin Wilson coefficients $C_{3-6}$ are counted as
${\cal O}(\alpha_s)$.  In order to match the precision of the experimental
value in \eq{eq:exp} one needs the yet unknown complete NNLO corrections
proportional to two factors of the current-current Wilson coefficients
$C_{1,2}$, while the contributions proportional to $C_{1,2}C_{3-6}$ are only
needed at NLO.  In Ref.~\cite{Asatrian:2020zxa} for the first time penguin
contributions have been considered beyond LO, presenting the terms
proportional to $C_{1,2}C_{3-6}\, \alpha_s N_f$.

In this paper we present the QCD corrections to all penguin
contributions proportional to the product of $C_{1,2}$ with one of
$C_{3-6}$ in an expansion in
\begin{eqnarray}
  z &=& \left(\frac{m_c^{\rm OS}}{m_b^{\rm OS}} \; \right)^2\,, \label{eq:defz}
\end{eqnarray}
where the superscript ``OS'' refers to the on-shell (or pole) scheme, i.e.\  
two-loop contributions of order ${\cal O}(\alpha_s)$. Thus this is a step
towards the completion of the NLO prediction of $\Gamma_{12}^q$, which is a
necessary preparation for NNLO. This calculation
is more convenient in the ``CMM'' operator basis of Ref.~\cite{Chetyrkin:1997gb},
which avoids problems in connection to $\gamma_5$.  We also adopt this
basis in the calculation presented in this paper. As a byproduct we reproduce
the NLO result for the contribution with two copies of $C_{1,2}$ of
Refs.~\cite{Beneke:1998sy,Ciuchini:2003ww,Beneke:2003az,Lenz:2006hd} (expanded
in $z$) after transforming the $\Delta B=1$ Wilson coefficients to the CMM
basis, which is a powerful check of our calculational set-up.

The paper is organised as follows: Sec.~\ref{sec:pr} introduces the
$\Delta B=1$ and $\Delta B=2$ operator bases employed by us,
Sec.~\ref{sec::calc} and Appendix~\ref{app::proj} present the methodology of
our calculation, Sec.~\ref{sec::res} contains the results, and we conclude in
Sec.~\ref{sec::concl}.

\section{Preliminaries\label{sec:pr}}
The effective $|\Delta B|=1$ weak Hamiltonian in the CMM operator
basis \cite{Chetyrkin:1997gb} reads:
\begin{eqnarray}
  \mathcal{H}_{\textrm{eff}}^{|\Delta B|=1} 
  &=&   \frac{4G_F}{\sqrt{2}}  \left[
      -\, \lambda^s_t \Big( \sum_{i=1}^6 C_i Q_i + C_8 Q_8 \Big) 
      - \lambda^s_u \sum_{i=1}^2 C_i (Q_i - Q_i^u) \right. \nn
  && \phantom{\frac{4G_F}{\sqrt{2}} \Big[}
      \left.
      +\, V_{us}^\ast V_{cb} \, \sum_{i=1}^2 C_i Q_i^{cu} 
      + V_{cs}^\ast V_{ub} \, \sum_{i=1}^2 C_i Q_i^{uc} 
      \right]
      + \mbox{h.c.}\,,
      \label{eq::HamDB1}
\end{eqnarray}
where $\lambda^s_a = V_{as}^\ast V_{ab}$, $a=u,c,t,$  contains the
CKM matrix elements and $\lambda_t=-\lambda_c-\lambda_u$.
For definiteness we specify to $b\to s$ decays relevant for
\bbms. The corresponding expressions for \bbmd\ are trivially found by
replacing $ V_{as}$ with  $V_{ad}$.
$G_F$ is the Fermi constant and the  dimension-six $\Delta B=1$ operators are given by
\begin{eqnarray}
  Q^u_1 &=& \bar{s}_L \gamma_{\mu} T^a u_L\;\bar{u}_L     \gamma^{\mu} T^a b_L\,,\nonumber \\
  Q^u_2 &=& \bar{s}_L \gamma_{\mu}     u_L\;\bar{u}_L     \gamma^{\mu}     b_L\,,\nonumber\\
  Q^{cu}_1 &=& \bar{s}_L \gamma_{\mu} T^a u_L\;\bar{c}_L     \gamma^{\mu} T^a b_L\,,\nonumber \\
  Q^{cu}_2 &=& \bar{s}_L \gamma_{\mu}     u_L\;\bar{c}_L     \gamma^{\mu}     b_L\,,\nonumber\\
  Q^{uc}_1 &=& \bar{s}_L \gamma_{\mu} T^a c_L\;\bar{u}_L     \gamma^{\mu} T^a b_L\,,\nonumber \\
  Q^{uc}_2 &=& \bar{s}_L \gamma_{\mu}     c_L\;\bar{u}_L     \gamma^{\mu}     b_L\,,\nonumber\\
  Q_1   &=& \bar{s}_L \gamma_{\mu} T^a c_L\;\bar{c}_L     \gamma^{\mu} T^a b_L\,,\nonumber\\
  Q_2   &=& \bar{s}_L \gamma_{\mu}     c_L\;\bar{c}_L     \gamma^{\mu}     b_L\,,\nonumber\\
  Q_3   &=& \bar{s}_L \gamma_{\mu}     b_L \sum_q \bar{q}\gamma^{\mu}     q\,,\nonumber\\
  Q_4   &=& \bar{s}_L \gamma_{\mu} T^a b_L \sum_q \bar{q}\gamma^{\mu} T^a q\,,\nonumber\\
  Q_5   &=& \bar{s}_L \gamma_{\mu_1}
            \gamma_{\mu_2}
            \gamma_{\mu_3}    b_L\sum_q \bar{q} \gamma^{\mu_1} 
            \gamma^{\mu_2}
            \gamma^{\mu_3}     q\,,\nonumber\\
  Q_6   &=& \bar{s}_L \gamma_{\mu_1}
            \gamma_{\mu_2}
            \gamma_{\mu_3} T^a b_L\sum_q \bar{q} \gamma^{\mu_1} 
            \gamma^{\mu_2}
            \gamma^{\mu_3} T^a q\,,\nonumber\\
  Q_8  &=&  \frac{g_s}{16\pi^2} m_b \, \bar{s}_L \sigma^{\mu \nu} T^a
           b_R \, G_{\mu\nu}^a\,,
           \label{operators}
\end{eqnarray}
where $q_L=P_Lq$ with $P_L=(1-\gamma_5)/2$.  $Q_1^{(u)}$ and
$Q_2^{(u)}$ are the current-current operators describing the
$W$-mediated tree-level decay of the $b$ quark including QCD effects.
$Q_3,\ldots,Q_8$ are four-quark penguin operators. We list the
operator $Q_8$ (with
  $\sigma^{\mu \nu} = i[\gamma^\mu,\gamma^\nu]/2$) for completeness; it
does not enter the calculations in this paper.  $g_s$ is the strong
coupling constant and $G^a_{\mu \nu}$ denotes the gluon field strength
tensor.  In Eq.~(\ref{operators}) the sum over $q$ runs over all five
quark fields $u,d,s,c$ or $b$.  For our calculation we also need the
following evanescent operators~\cite{Chetyrkin:1997gb}
\begin{eqnarray}
  E_1[Q_1] &=& 
                \bar{s}_L \gamma^{\mu_1} \gamma^{\mu_2} \gamma^{\mu_3}
               T^a c \;
                \bar{c} \gamma_{\mu_1} \gamma_{\mu_2} \gamma_{\mu_3}  T^a
                b_L - 16 Q_1\,,\nonumber\\ 
  E_1[Q_2] &=& 
                \bar{s}_L \gamma^{\mu_1} \gamma^{\mu_2} \gamma^{\mu_3}
                c_i\;\bar{c}_j \gamma_{\mu_1} \gamma_{\mu_2}
                \gamma_{\mu_3} b_L - 16 Q_2\,,\nonumber\\ 
  E_1[Q_5] &=& 
                \bar{s}_L \gamma^{\mu_1} \gamma^{\mu_2} \gamma^{\mu_3}
                \gamma^{\mu_4} \gamma^{\mu_5} b_L \sum_q \bar{q}
                \gamma_{\mu_1} \gamma_{\mu_2} \gamma_{\mu_3} \gamma_{\mu_4}
                \gamma_{\mu_5} q_j \, - \, 20 Q_5 \, +\,  64 Q_3\,,\nonumber\\ 
  E_1[Q_6] &=& 
                \bar{s}_L \gamma^{\mu_1} \gamma^{\mu_2} \gamma^{\mu_3}
                \gamma^{\mu_4} \gamma^{\mu_5} T^a b_L \sum_q
                \bar{q} \gamma_{\mu_1} \gamma_{\mu_2} \gamma_{\mu_3}
                \gamma_{\mu_4} \gamma_{\mu_5} T^a q \,-\, 20 Q_6 \,+\, 64 Q_4 
                \,
\label{evan_operators}
\end{eqnarray}
and the counterparts of $E_1[Q_{1,2}]$ with one or both $c$ replaced by $u$. The
$\Delta B=1$ operators in \eqsand{operators}{evan_operators} destroy a
$b$ and $\bar s$ quark while creating a $\bar b$ and $s$ quark and
thereby describe the transition of a $\bar B_s \sim b\bar s$ into a
$B_s \sim \bar b s$ meson. The corresponding Feynman diagrams have
incoming $b$ quark and outgoing $s$ quark lines.

Using the Hamiltonian in Eq.~(\ref{eq::HamDB1}) the width difference
$\Delta\Gamma\approx 2 |\Gamma_{12}|$ is obtained from
\begin{eqnarray}
  \Gamma_{12}^s &=& \frac{1}{2 M_{B_s}} \,\mbox{Abs}\langle B_s|i\int{\rm d}^4 x \,\, T\,\,
                    {\cal H}_{\rm eff}^{\Delta B=1}(x)
                    {\cal H}_{\rm eff}^{\Delta B=1}(0)
                    |\bar{B}_s\rangle\,, \label{eq:ot}
\end{eqnarray}
where ``Abs'' stands for the absorptive part and $T$ is the time ordering
operator. $\Gamma_{12}^s$ encodes the information of the inclusive
decay rate into final states common to $B_s$ and $\bar B_s$ and
\eq{eq:ot} employs the optical theorem to relate $\Gamma_{12}^s$ to
the $\bar B_s \to B_s$ forward scattering amplitude.

It is convenient to decompose $\Gamma_{12}^s$ as~\cite{Beneke:1998sy}
\begin{eqnarray}
  \Gamma_{12}^q &=& - (\lambda_c^q)^2\Gamma^{cc}_{12} 
                  - 2\lambda_c^q\lambda_u^q \Gamma_{12}^{uc} 
                  - (\lambda_u^q)^2\Gamma^{uu}_{12} 
                  \,,
                    \label{eq::Gam12}
\end{eqnarray}
where in the practical calculation the quantities $\Gamma_{12}^{ab}$ are
considered.

The Heavy Quark Expansion (HQE) allows us to express the quantities
$\Gamma_{12}^{ab}$ in Eq.~(\ref{eq::Gam12}) in terms of
infrared-safe perturbative coefficients and hadronic matrix elements
of $\Delta B=2$ operators. To leading power in $1/m_b$ one only
needs two $\Delta B=2$ operators, which are conveniently chosen as
\begin{eqnarray}
  Q &=& \bar{s}_i \gamma^\mu \,(1-\gamma^5)\, b_i \; \bar{s}_j \gamma_\mu
        \,(1-\gamma^5)\, b_j\,, \nonumber\\
  \widetilde{Q}_S &=& \bar{s}_i \,(1-\gamma^5)\, b_j\; \bar{s}_j \,(1-\gamma^5)\,
                  b_i\, \label{eq::opDB2}
\end{eqnarray}
with colour indices $i,j$. At intermediate steps of the calculation one also encounters
\begin{eqnarray}
\widetilde{Q} &=& \bar{s}_i \gamma^\mu \,(1-\gamma^5)\, b_j \; \bar{s}_j \gamma_\mu
                 \,(1-\gamma^5)\, b_i\,,\nonumber\\
  Q_S &=& \bar{s}_i \,(1-\gamma^5)\, b_i \; \bar{s}_j \,(1-\gamma^5)\,
         b_j\,,
          \label{eq:defqs}
\end{eqnarray}
and operators with more than two Dirac matrices on both quark
lines. $Q_S$ can be traded for $Q$, $\widetilde Q_S$, and an operator
$R_0$ describing $1/m_b$-suppressed contributions to $\Gamma_{12}^s$ \cite{Beneke:1996gn},
\begin{eqnarray}
 Q_S &=& -\widetilde Q_S - \frac{1}{2}
                                  Q + R_0. \label{eq:defr0}
\end{eqnarray}  
By subtracting judiciously constructed linear combinations of $Q$ and
$\widetilde Q_S$, all additional operators entering the calculation are evanescent, meaning
that they vanish in $D=4$ dimensions. We choose
\cite{Beneke:1998sy,Gorbahn:2009pp}

\begin{eqnarray}
  E_1^{(1)} &=& \widetilde{Q} - Q\,, \nonumber\\
  E_2^{(1)} &=& \bar{b}_i \gamma^\mu \gamma^\nu \gamma^\rho \,P_L\,
                s_j
          \bar{b}_j \gamma_\mu \gamma_\nu \gamma_\rho \,P_L\, s_i - (16 -4 \epsilon) \widetilde{Q}\,, \nonumber\\
  E_3^{(1)} &=& \bar{b}_i \gamma^\mu \gamma^\nu \gamma^\rho \,P_L\,
                s_i  \bar{b}_j \gamma_\mu \gamma_\nu \gamma_\rho \,P_L\, s_j - (16 -4 \epsilon) Q\,, \nonumber\\
  E_4^{(1)} &=& \bar{b}_i \gamma^\mu \gamma^\nu \,P_L\, s_j
                \bar{b}_j \gamma_\nu \gamma_\mu \,P_L\, s_i  + (8 - 8 \epsilon) Q_s\,, \nonumber\\
  E_5^{(1)} &=& \bar{b}_i \gamma^\mu \gamma^\nu \,P_L\, s_i
                \bar{b}_j \gamma_\nu \gamma_\mu \,P_L\, s_j  + (8 - 8 \epsilon) \widetilde{Q}_s\,,
\label{eq:E1}
\end{eqnarray}
with the usual $\epsilon=(4-D)/2$ of dimensional regularisation. The operators
on the RHS are understood to be expressed in terms of the minimal physical
basis $Q\,$,$\widetilde Q_S$, e.g.\ $\widetilde{Q}$ is to be read as
$Q+E_1^{(1)}$ in $E_2^{(1)}$.  The choice of the ${\cal O}(\epsilon)$ terms in
the coefficients affect the expressions of the renormalised coefficients
$H^{ab}\,$, $\widetilde H^{ab}_S$ of $Q\,$,$\widetilde Q_S$
\cite{Herrlich:1994kh}. That is, their specification is part of the definition
of the renormalisation scheme of the operators (along with the
$\overline{\rm MS}$ prescription and the use of anticommuting $\gamma_5$). Our
definitions in \eq{eq:E1} ensure that the coefficients do not depend on the
Fierz arrangement \cite{Herrlich:1994kh,Gorbahn:2009pp}, i.e.\ a
four-dimensional Fierz transformation of $Q$, $\widetilde Q_S$ does not change
$C\,$ and $\widetilde C_S$.

It is thus possible to write
$\Gamma_{12}^{ab}$ in \eq{eq::Gam12} as
\begin{eqnarray}
  \Gamma_{12}^{ab} 
  &=& \frac{G_F^2m_b^2}{24\pi M_{B_s}} \left[ 
      H^{ab}(z)   \langle B_s|Q|\bar{B}_s \rangle
      + \widetilde{H}^{ab}_S(z)  \langle B_s|\widetilde{Q}_S|\bar{B}_s \rangle
      \right]
      + \ldots \,
      \label{eq::Gam^ab}
\end{eqnarray}
with $z$ defined in \eq{eq:defz}.
The ellipses denote higher-order terms in $\Lambda_{\rm QCD}/m_b$.
The matching coefficients $H^{ab}$ and $\widetilde{H}_S^{ab}$ are related to the
functions $G^{ab}$ and $G_S^{ab}$ defined in Refs.~\cite{Beneke:1998sy}
via (see, e.g., Eq.~(21) of Ref.~\cite{Lenz:2006hd})
\begin{eqnarray}
  H^{ab} &=& G^{ab} + \frac{\alpha_2}{2} G_S^{ab}\,,\nonumber\\
  \widetilde{H}_S^{ab} &=& G_S^{ab} \alpha_1\,,
\end{eqnarray}
with 
\begin{eqnarray}
  \alpha_1 &=& 1 + \frac{\alpha_s(\mu_2)}{4\pi} C_F 
               \left(6 + 12\log\frac{\mu_2}{m_b}\right)\,,\nonumber\\
  \alpha_2 &=& 1 + \frac{\alpha_s(\mu_2)}{4\pi} C_F 
               \left(\frac{13}{2} + 6\log\frac{\mu_2}{m_b}\right)\,,
\end{eqnarray}
where $C_F=(N_c^2-1)/(2N_c)$ with $N_c=3$ denoting the number of colours.
We decompose $H^{ab}(z)$ and $\widetilde{H}_S^{ab}(z)$ as follows
\begin{eqnarray}
  H{}^{ab}(z) &=& \F{ab} (z)+ \PP{ab}(z)  +\PPP{ab}(z) \,, \nonumber\\
  \widetilde{H}_S^{ab}(z) &=& \FS{ab}(z)+ \PS{ab}(z) +\PPS{ab}(z)\,,
             \label{eq::G}
\end{eqnarray}
where the superscript ``(c)'' denotes the contributions with two
current-current operators $Q_{1,2}$, while ``(cp)'' refers to those with one
operator $Q_{1,2}$ and one penguin operator $Q_{3-6}$ and ``(p)'' labels the
terms involving two penguin operators.  The functions $\PP{ab}(z)$ and
$\PS{ab}(z)$ are the main focus of this paper.

%- }}}
%- {{{ Calculation:

\section{\label{sec::calc}Calculation}
The Wilson coefficients $H^{ab}(z)$ and $\widetilde{H}_S^{ab}(z)$ encode the
short-distance physics and are independent of the external states in the
matrix elements in \eqsand{eq:ot}{eq::Gam^ab}. Thus one may replace the mesons
by free quarks, i.e.\ calculate the forward-scattering amplitude
\begin{eqnarray}
  b + \bar{s} \to \bar{b} + s   \no
\end{eqnarray}
in perturbation theory and apply the optical theorem in order to extract the
desired absorptive part.  By equating \eq{eq:ot} with \eq{eq::Gam^ab} one
determines $H^{ab}(z)$ and $\widetilde{H}_S^{ab}(z)$.  The infrared
singularities present in both sides of this matching equation factorise, which
makes the desired coefficients meaningful infrared-safe perturbative
quantities.  The external quarks are on-shell, i.e.\ we have $p_b^2=m_b^2$ and
may choose $p_s=0$ since we use $m_s=0$ and terms proportional to
$p_b\cdot p_s$ match onto power-suppressed matrix elements. Thus we must
evaluate two-point loop integrals with external momentum $q^2=m_b^2$.  In our
calculation we regulate the infrared divergences with a gluon mass which
introduces a further mass scale, $m_g$. We introduce the gluon propagator as
\begin{eqnarray}
  \frac{i \delta^{ab} \left (g^{\mu \nu} + \xi \frac{p^\mu p^\nu}{-p^2-i0} \right) }
  {m_g^2 - p^2 - i0} 
  \label{eq:gluonprop}
  \,.
\end{eqnarray}
It is possible to expand the Feynman integrals for $m_g\ll m_b$. We perform
this expansion at the level of the master integrals as described below. We
further employ an arbitrary QCD gauge parameter $\xi$ and use its cancellation
as a check of our calculation.

In the following we describe our methodology for the dominant contribution
encoded in $H^{cc}(z)$ and $\widetilde{H}_S^{cc}(z)$. The calculational steps
for the CKM-suppressed contributions involving $H^{uc}(z)$ and
$\widetilde{H}_S^{uc}(z)$ are the same.  Our practical calculation proceeds as
follows: We consider the bilocal matrix elements
\begin{eqnarray}
  \mbox{Abs} \langle \, i \!\int \! {\rm d}^4 x \;T O_i(x) O_j(0) \, \rangle
  \,,
  \label{eq::biME}
\end{eqnarray}
where $O_i$ and $O_j$ are operators from Eqs.~(\ref{operators})
and~(\ref{evan_operators}). At one-loop order we have to consider the cases
$O_i,O_j \in \{Q_1,\ldots,Q_6\}$ and $O_i \in \{Q_1,\ldots,Q_6\}$,
$O_j \in \{E_1^{(1)},\ldots,E_4^{(1)}\}$.  The matrix elements with evanescent
operators enter via the renormalisation procedure. One may formulate this
procedure in terms of either bare and renormalised Wilson coefficients or bare
and renormalised operators. With the former choice we have
\begin{eqnarray}
  ( C_1,\ldots, C_6,C_{E_1^{(1)}},\ldots,C_{E_4^{(1)}} )^{\rm bare}
  &=&
  ( C_1,\ldots, C_6, * ) 
  \left(
  \begin{array}{cc} 
    Z_{QQ} & Z_{QE} \\
    * & *
  \end{array}
        \right)\,
        \label{eq::C_ren}
\end{eqnarray}
where $Z_{QQ}$ and $Z_{QE}$ are $6\times6$ and $6\times4$ matrices,
respectively. They can be extracted from Ref.~\cite{Gambino:2003zm}.
The entries in Eq.~(\ref{eq::C_ren}) represented by a $*$ are
irrelevant for our calculation. The UV poles contained in  $Z_{QQ}$
and $Z_{QE}$ force us to include
${\cal O}(\epsilon)$ terms in the one-loop matrix elements
$\langle \, i \!\int \! {\rm d}^4 x \, T O_i(x) O_j(0) \,
\rangle^{(0)}$ multiplied by $Z_{QQ}$,$Z_{QE}$.

At two loops we compute $\langle \, i \! \int \! {\rm d}^4 x \, T O_i(x) O_j(0)
\, \rangle $ for $O_i,O_j \in \{Q_1,Q_2\}$, transform the
result to the traditional operator basis
\cite{Buras:1992tc,Buchalla:1995vs}, and compare to the
literature~\cite{Beneke:1998sy} in order to have a
non-trivial cross check for the implementation of the CMM operator basis.
New results are obtained for $\langle \, i \! \int \! {\rm d}^4 x \, T
Q_{1-2}(x) Q_{3-6}(0) \, \rangle$.

For our calculation we use a well-tested program chain including {\tt
  qgraf}~\cite{Nogueira:1991ex} for the generation of the amplitudes, {\tt
  q2e} and {\tt exp}~\cite{Harlander:1997zb,Seidensticker:1999bb} for the
identification of the integral families and {\tt FORM}~\cite{Kuipers:2012rf}
for the algebraic manipulations and the traces of the $\gamma$ matrices.  As
an alternative to {\tt q2e} we also use the program {\tt tapir}~\cite{tapir}
which automatically generates {\tt FORM} code, in which scalar products in the
numerator are re-written in denominator factors and relations implementing a
partial fraction decomposition are applied, if necessary. Furthermore, the
input files for {\tt FIRE}~\cite{Smirnov:2019qkx} are automatically generated.
The Feynman rules involving the $\Delta B=1$ and
$\Delta B=2$ operators have been obtained with the help of {\tt
  FeynRules}~\cite{Alloul:2013bka} and {\tt
  FeynCalc}~\cite{Shtabovenko:2016sxi,Shtabovenko:2020gxv}.

\begin{figure}[t]
  \begin{center}
    \begin{tabular}{cccc}
      \multicolumn{4}{c}{
      \includegraphics[width=0.18\textwidth]{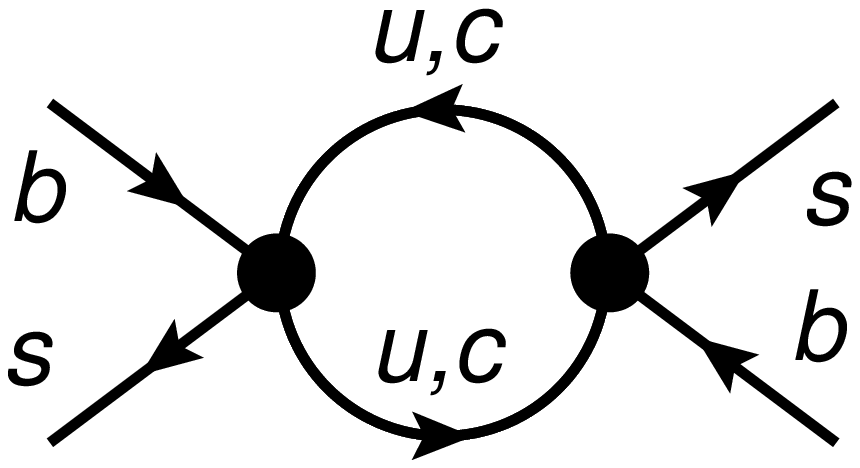}
      }\\
      \multicolumn{4}{c}{(a)}\\
  \includegraphics[width=0.18\textwidth]{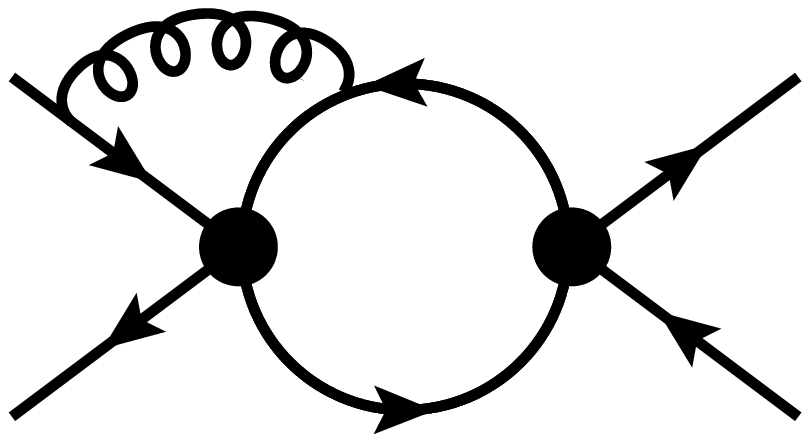} &
  \includegraphics[width=0.25\textwidth]{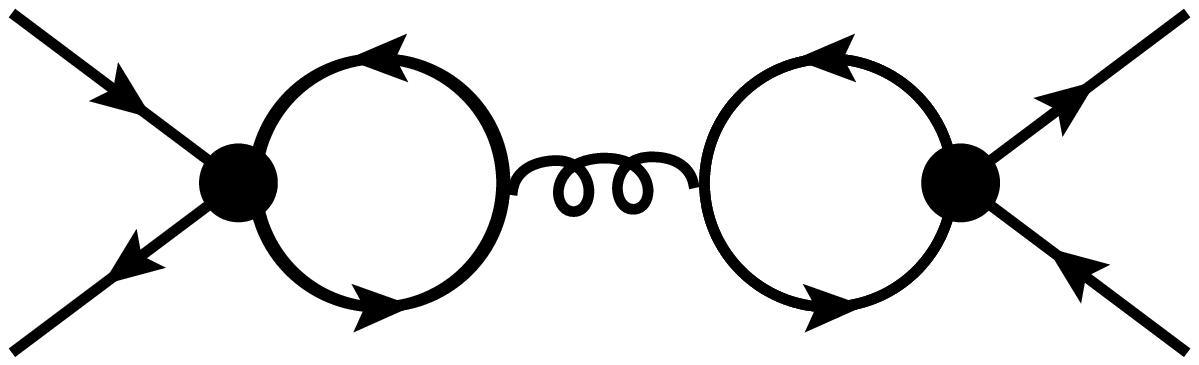} &
  \includegraphics[width=0.18\textwidth]{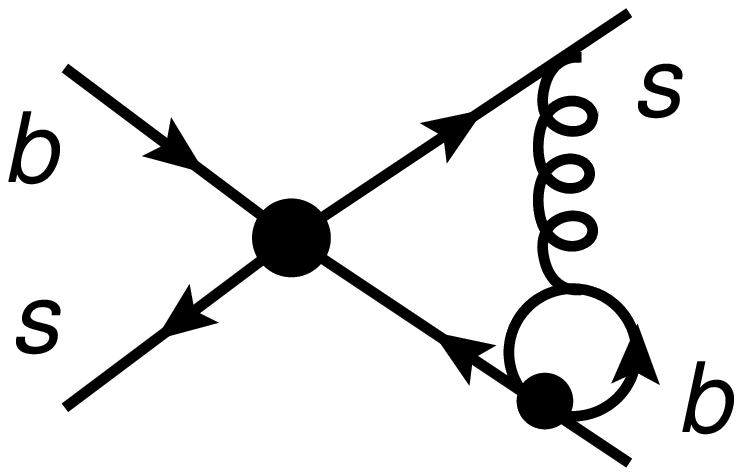} &
  \includegraphics[width=0.25\textwidth]{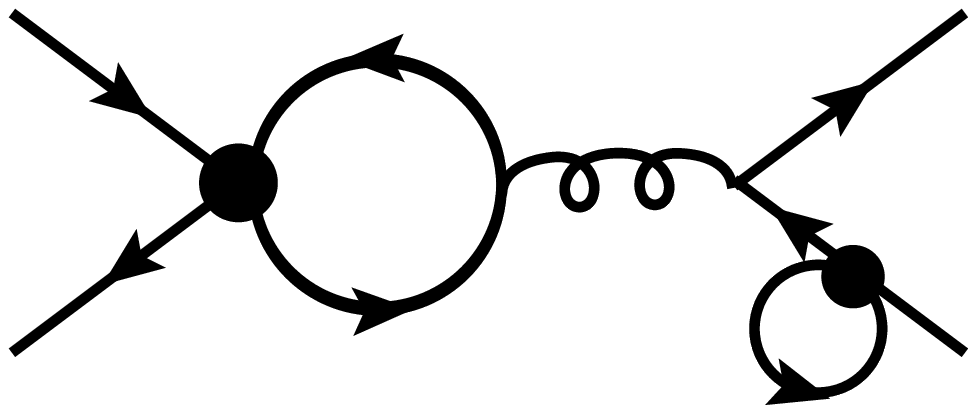} \\
  (b) & (c) & (d) & (e)
    \end{tabular}
  \end{center}
  \caption{\label{fig::sample}Sample Feynman diagrams contribution to the
    process $b + \bar{s} \to \bar{b} + s$ with $\Delta B=1$ operators.  The
    latter are marked by a blob. In (a), (b) and (c) both operators can be
    from the set $\{Q_1,\ldots, Q_6\}$ whereas in (d) and (e) one of the operators has
    to be from the set $\{Q_3,\ldots, Q_6\}$.}
\end{figure}

At one-loop order only the type of diagrams shown in
Fig.~\ref{fig::sample}(a) contribute. At two loops we can distinguish four
different classes of Feynman diagrams, see also Fig.~\ref{fig::sample}.
Figures~\ref{fig::sample}(b) and (c) show the type of diagrams which
contribute to the matrix element $\langle \, i \! \int \! {\rm
    d}^4 x \,T Q_{1,2}(x) Q_{1,2}(0) \, \rangle $. These
topologies are also present if one of the operators is replaced by a penguin
operator.  Note that in Fig.~\ref{fig::sample}(c) one of the closed quark
loops contains charm or up quarks whereas the other may contain all five active
flavours.  In Fig.~\ref{fig::sample}(d) and (e) we show sample diagrams which
require the presence of a penguin operator. In Fig.~\ref{fig::sample}(d) it is
the left operator whereas in (e) it is the one on the external quark
line.

We have implemented two approaches for the manipulation of the fermion spinor
lines. In the first approach we concentrate on tensor integrals and various
manipulations of Dirac structures.  We use {\tt
  FeynCalc}~\cite{Mertig:1990an,Shtabovenko:2016sxi,Shtabovenko:2020gxv} 
together with {\tt Fermat}~\cite{fermat} to obtain formulae
for tensor reduction which we then implement in {\tt FORM}.  
To this end the tensor reduction algorithm of {\tt FeynCalc} was improved
using ideas from~\cite{Pak:2011xt}.
In the second
approach we construct projectors to all Dirac structures.  This has the
advantage that we can take traces and afterwards only scalar expressions have
to be manipulated. More details can be found in Appendix~\ref{app::proj}.

At this point a comment concerning the expansion in $m_c$ is in order.  Since
we restrict ourselves to quadratic terms in $m_c$, i.e.\ linear
terms in $z$, all loop integrals with both  
bottom and charm
quark lines present in the same loop can be naively Taylor-expanded in
$m_c$ before performing the
loop integrations.\footnote{There is, however, a $z\log z$ term in
    one-loop diagrams with a charm mass counterterm. This term does not affect
    the expansion of the unrenormalised two-loop integrals in $z$
    and, moreover, is absent once the result is expressed in terms
    of $\bar z =m_c(m_b)/m_b(m_b)$ \cite{Beneke:2002rj}.} All such
diagrams are contained in the class which is represented by
Fig.~\ref{fig::sample}(a) and~(b). In all other cases we can apply the so-called
large-momentum expansion~\cite{Smirnov:2006ry} as implemented in {\tt
  exp}~\cite{Harlander:1997zb,Seidensticker:1999bb}. However,
our explicit calculation shows that up to order $z$ 
indeed a naive expansion in $m_c$ is sufficient.

For the reduction to master integrals we use {\tt FIRE}~\cite{Smirnov:2019qkx}
and {\tt LiteRed}~\cite{Lee:2012cn,Lee:2013mka}. For all infrared
contributions the reduction is performed for general gluon mass $m_g$.
Afterwards we consider the limit of small $m_g$ and perform an asymptotic
expansion~\cite{Smirnov:2006ry} for $m_g\ll m_b$ at the level of the master
integrals. We have performed numerical cross-checks of the expansions with the
help of {\tt FIESTA}~\cite{Smirnov:2015mct}.  After the asymptotic expansion
we have to compute single-scale one- and two-loop integrals, most of
which  are available in the literature (see, e.g., Ref.~\cite{Smirnov:2006ry}).
The remaining ones are straightforward to compute.

We multiply the matrix element on both sides of the matching
equation  with $Z_\psi^2$, where
$Z_\psi$ is the quark field renormalisation constant in the
$\ov{\rm MS}$ scheme. This renders both expressions
UV-finite. Note, that they still depend on the gauge
parameter which is due to the gluon mass used as infrared regulator.
For the renormalisation of the charm quark mass we use both the
$\overline{\rm MS}$ and on-shell scheme, see also Section~\ref{sec::res}.
No renormalisation of the bottom or strange quark mass is
needed since in the considered order there are no corresponding self-energy
diagrams.

\begin{figure}[t]
  \begin{center}
      \includegraphics[width=0.9\textwidth]{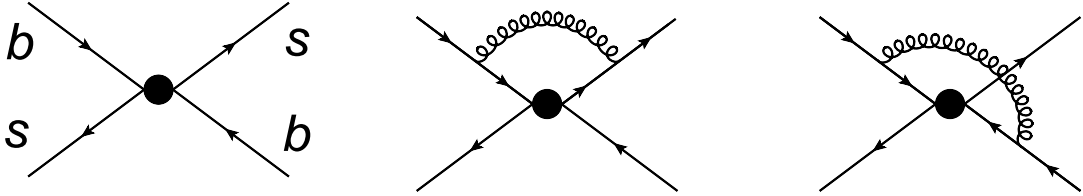}
  \end{center}
  \caption{\label{fig::diasDB2}Sample Feynman diagrams with $\Delta B=2$
    operators.}
\end{figure}
For the $\Delta B=2$ theory we calculate one-loop QCD corrections for the
matrix elements of the minimal operator basis in \eq{eq::opDB2}.  Sample
Feynman diagrams, which have to be considered at NLO, are shown in
Fig.~\ref{fig::diasDB2}.  The results of the matrix elements in both the
$\Delta B=1$ and $\Delta B=2$ theories can be expressed as a linear
combination of the tree-level matrix elements of $Q$, $\widetilde{Q}_S$, $R_0$
and the unphysical operators in \eq{eq:E1}. Since both results are UV-finite
we can take the limit $\epsilon\to 0$ and then read off the desired NLO
corrections to the $\Delta B=2$ Wilson coefficients $H^{ab}$ and
$\widetilde H_S^{ab}$.  We observe that the infrared regulator $m_g$ and the
gauge parameter cancel from these coefficients, providing a non-trivial check
of the calculation. $H^{ab}$ and $\widetilde H_S^{ab}$ depend on the
renormalisation scales $\mu_1$ and $\mu_2$, at which the renormalised
operators are defined in the $\Delta B=1$ and $\Delta B=2$ theories, respectively.  The
$\mu_1$-dependence of $H^{ab}$ and $\widetilde H_S^{ab}$ diminishes
order-by-order in perturbation theory and is commonly used as a means to
estimate the accuracy of the truncated perturbative series. The
$\mu_2$-dependence cancels in the matching procedure of the perturbative
$\Delta B=2$ matrix elements with their non-perturbative counterparts.

%- }}}
%- {{{ Analytical and numerical results:

\section{\label{sec::res}Analytical and numerical results}

In the following we discuss the results for the matching coefficients $H^{ab}$
and $\widetilde{H}^{ab}_S(z)$ introduced in Eq.~(\ref{eq::Gam^ab}).

We start with the analytic expressions  for the
penguin contributions $\PP{ab}(z)$ and $\PS{ab}(z)$ (for $ab = uu,uc$
and $cc$) introduced in Eq.~(\ref{eq::G}).  It is convenient to
decompose the $\Delta B=2$ matching coefficients in terms of the
$\Delta B=1$ coefficients $C_i$ of the $|\Delta B|=1$ Hamiltonian in
\eq{eq::HamDB1}:
\begin{eqnarray}
  \F{ab} (z) &=& \sum_{i,j=1}^2 C_i C_j \, p_{ij}^{ab} (z)\,,
                 \nonumber\\
  \FS{ab} (z) &=&  \sum_{i,j=1}^2 C_i C_j \, p_{ij}^{S,ab} (z)\,, \nonumber\\
  \PP{ab} (z) &=& \sum_{i=3,\ldots 6,8} C_{i} \lt[ C_1 p_{1i}^{ab} (z) +C_2 p_{2i}^{ab} (z) \rt]
                  \,,
                  \nonumber\\
  \PS{ab} (z) &=& \sum_{i=3,\ldots 6,8} C_{i} \lt[ C_1 p_{1i}^{S,ab} (z) +C_2 p_{2i}^{S,ab} (z)\rt] 
                  \,, \nonumber\\
  \PPP{ab} (z)  &=&   \sum_{i,j=3,\ldots 6,8} C_i C_{j}\,p_{ij}^{ab} (z)
                    \,, \nonumber\\ 
  \PPS{ab} (z)  &=&
                    \sum_{i,j=3,\ldots 6,8} C_i C_{j} \,p_{ij}^{S,ab} (z)
                    \,.
                    \label{eq::Hp_HSp}
\end{eqnarray}
We furthermore introduce the perturbative expansion as
\begin{eqnarray}
  p_{ij}^{ab}(z) = p_{ij}^{ab,(0)}(z) + \frac{\alpha_s(\mu_1)}{4\pi}
  p_{ij}^{ab,(1)} (z) + {\cal O}(\alpha_s^2)\,, \label{eq:pp}
\end{eqnarray}
(and analogously for the other coefficients) where $p_{ij}^{ab,(0)}$ refers to
one-loop and $p_{ij}^{ab,(1)}$ to two-loop contributions.  In this paper the strong coupling
constant is defined with five active quark flavours at the renormalisation
scale $\mu_1$, i.e.\ we have $\alpha_s\equiv\alpha_s^{(5)}(\mu_1)$.
For later convenience we introduce the squared ratio of the charm and bottom
quark masses as
\begin{eqnarray}
  z &=&
  \left(\frac{m_c^{\rm OS}}{m_b^{\rm OS}} \; \right)^2 
  \;=\;  
  \left(\frac{\overline{m}_c(m_c)}{\overline{m}_b(m_b)} \; \right)^2 
  \,+\, {\cal O} \lt( \alpha_s^2\rt)\,, 
  \qquad
  \qquad
  \bar z \; =
  \;
  \left(\frac{\overline{m}_c(m_b)}{\overline{m}_b(m_b)} \; \right)^2
  \,,
  \label{eq:defzbar}
\end{eqnarray}
with the $\ov{\rm MS}$ masses $\ov m_q$ and the pole (on-shell) masses
$m_q^{\rm OS}$. While it is easier to employ on-shell masses in the
calculation, their poor definition (especially of $m_c^{\rm OS}$) make them
unsuited for numerical evaluations and we will always use $\ov{\rm MS}$ values
as inputs.

The one-loop coefficients $p_{ij}^{ab,(0)}$, $p_{ij}^{S,ab,(0)}$ can be extracted from
Ref.~\cite{Beneke:1996gn}, where the full $m_c$ dependence has been taken into
account, by transforming the result to the operator bases used in
this paper. We can reproduce these results in an expansion in $z$ including the
linear terms. Note that $p_{i8}^{ab,(0)}$, $p_{i8}^{S,ab,(0)}$ and
$p_{88}^{ab,(1)}$, $p_{88}^{S,ab,(1)}$ vanish. For $ab=cc$ the non-zero
LO coefficients are
\begin{align}
	p^{cc,(0)}_{13}(z) &= \sqrt{1-4 z} \left(\frac{4}{3}+\frac{8 z}{3}\right), &\qquad
	p^{cc,(0)}_{14}(z) &= \sqrt{1-4 z} \left(-\frac{5}{36}-\frac{5 z}{18}\right), \nonumber\\
	p^{cc,(0)}_{15}(z) &= \sqrt{1-4 z} \left(\frac{64}{3}-\frac{160 z}{3}\right), &\qquad
	p^{cc,(0)}_{16}(z) &= \sqrt{1-4 z} \left(-\frac{20}{9}-\frac{4 z}{9}\right), \nonumber\\
	p^{cc,(0)}_{23}(z) &= \sqrt{1-4 z} (1+2 z), &\qquad
	p^{cc,(0)}_{24}(z) &= \sqrt{1-4 z} \left(\frac{5}{6}+\frac{5 z}{3}\right), \nonumber\\
	p^{cc,(0)}_{25}(z) &= \sqrt{1-4 z} (16-40 z) , &\qquad
	p^{cc,(0)}_{26}(z) &= \sqrt{1-4 z} \left(\frac{40}{3}+\frac{8 z}{3}\right),
\end{align}
as well as
\begin{align}
	p^{S,cc,(0)}_{13}(z) &= \sqrt{1-4 z} \left(-\frac{8}{3}-\frac{16 z}{3}\right) , &\qquad
	p^{S,cc,(0)}_{14}(z) &= \sqrt{1-4 z} \left(-\frac{2}{9}-\frac{4 z}{9}\right), \nonumber\\
	p^{S,cc,(0)}_{15}(z) &= \sqrt{1-4 z} \left(-\frac{128}{3}-\frac{256 z}{3}\right) , &\qquad
	p^{S,cc,(0)}_{16}(z) &= \sqrt{1-4 z} \left(-\frac{32}{9}-\frac{64 z}{9}\right) , \nonumber\\
	p^{S,cc,(0)}_{23}(z) &= \sqrt{1-4 z} (-2-4 z) , &\qquad
	p^{S,cc,(0)}_{24}(z) &= \sqrt{1-4 z} \left(\frac{4}{3}+\frac{8 z}{3}\right), \nonumber\\
	p^{S,cc,(0)}_{25}(z) &= \sqrt{1-4 z} (-32-64 z) , &\qquad
	p^{S,cc,(0)}_{26}(z) &= \sqrt{1-4 z} \left(\frac{64}{3}+\frac{128 z}{3}\right).
\end{align}

The  two-loop coefficients $p_{ij}^{ab,(1)}$  are new
and are given by
\begin{align}
p^{cc,(1)}_{13}(z) &= \left(\frac{47}{18}-4 z\right) \logOne+\frac{56}{9} \logTwo +\frac{320 z}{9}+\frac{1523}{108}-\frac{5 \pi }{18 \sqrt{3}}, \nonumber\\
%%%%%%%%%%%%%%%%%%%%%%%%%%%
p^{cc,(1)}_{14}(z) &= \left(-\frac{371}{108}+\frac{5 N_H}{54}+\frac{5 N_L}{27}+\frac{5 N_V}{27}+\frac{59 z}{3}\right) \logOne+\frac{1}{54} \logTwo \nonumber  \\
& +\left(\frac{4265}{108}+\frac{5 N_L}{9}+\frac{10 N_V}{9}+\frac{5 \pi ^2}{9}\right) z \nonumber \\
& -\frac{1649}{162}+\frac{35 N_L}{162}+\frac{35 N_V}{162}+\frac{5 \pi }{108 \sqrt{3}}+\frac{5 \pi ^2}{18}+N_H \left(\frac{85}{162}-\frac{5 \pi }{18 \sqrt{3}}\right), \nonumber\\
%%%%%%%%%%%%%%%%%%%%%%%%%%%
p^{cc,(1)}_{15}(z) &= \left(\frac{376}{9}-136 z\right) \logOne+\left(\frac{896}{9}-192 z\right) \logTwo  +z \left(-\frac{16408}{9}-768 \log (z)\right) \nonumber \\
& +318-\frac{40 \pi }{9 \sqrt{3}}, \nonumber\\
%%%%%%%%%%%%%%%%%%%%%%%%%%%
p^{cc,(1)}_{16}(z) &= \left(-\frac{1484}{27}+\frac{25 N_H}{27}+\frac{50 N_L}{27}+\frac{50 N_V}{27}+\frac{764 z}{3}\right) \logOne+\left(\frac{8}{27}+8 z\right) \logTwo \nonumber \\
& +z \left(\frac{22100}{27}+\frac{50 N_L}{9}+\frac{100 N_V}{9}+\frac{8 \pi ^2}{9}+32 \log (z)\right) \nonumber \\
& -\frac{4543}{27}+\frac{130 N_L}{81}+\frac{130 N_V}{81}+\frac{20 \pi }{27 \sqrt{3}}+\frac{40 \pi ^2}{9}+N_H \left(\frac{380}{81}-\frac{25 \pi }{9 \sqrt{3}}\right),\nonumber\\ 
%%%%%%%%%%%%%%%%%%%%%%%%%%%
p^{cc,(1)}_{23}(z) &= \left(-\frac{47}{3}+24 z\right) \logOne+\frac{14}{3} \logTwo+\frac{170 z}{3}+\left(-\frac{677}{18}+\frac{5 \pi }{3 \sqrt{3}}\right), \nonumber\\
%%%%%%%%%%%%%%%%%%%%%%%%%%%
p^{cc,(1)}_{24}(z) &= \left(\frac{10}{9}-\frac{5 N_H}{9}-\frac{10 N_L}{9}-\frac{10 N_V}{9}+26 z\right) \logOne-\frac{1}{9} \logTwo \nonumber  \\
& +\left(\frac{1729}{18}-\frac{10 N_L}{3}-\frac{20 N_V}{3}-\frac{10 \pi ^2}{3}\right) z \nonumber \\
& + \frac{137}{27}-\frac{35 N_L}{27}-\frac{35 N_V}{27}-\frac{5 \pi }{18 \sqrt{3}}-\frac{5 \pi ^2}{3}+N_H \left(-\frac{85}{27}+\frac{5 \pi }{3 \sqrt{3}}\right),\nonumber\\
%%%%%%%%%%%%%%%%%%%%%%%%%%%
p^{cc,(1)}_{25}(z) &= \left(-\frac{752}{3}+816 z\right) \logOne+\left(\frac{224}{3}-144 z\right) \logTwo +z \left(\frac{3656}{3}-576 \log (z)\right) \nonumber \\
& -580+\frac{80 \pi }{3 \sqrt{3}}, \nonumber\\
%%%%%%%%%%%%%%%%%%%%%%%%%%%
p^{cc,(1)}_{26}(z) &= \left(\frac{160}{9}-\frac{50 N_H}{9}-\frac{100 N_L}{9}-\frac{100 N_V}{9}+128 z\right) \logOne+\left(-\frac{16}{9}-48 z\right) \logTwo  \nonumber \\
& +z \left(\frac{7640}{9}-\frac{100 N_L}{3}-\frac{200 N_V}{3}-\frac{16 \pi ^2}{3}-192 \log (z)\right) \nonumber \\
& + \frac{158}{9}-\frac{260 N_L}{27}-\frac{260 N_V}{27}-\frac{40 \pi }{9 \sqrt{3}}-\frac{80 \pi ^2}{3}+N_H \left(-\frac{760}{27}+\frac{50 \pi }{3 \sqrt{3}}\right)
\label{eq::p_ij^(1)}
\end{align}
and
\begin{align}
p^{S,cc,(1)}_{13}(z) &= -\frac{4}{3} \logOne-\frac{64}{9} \logTwo -\frac{1720 \
	z}{9} -\frac{130}{27}-\frac{4 \pi }{9 \sqrt{3}}, \nonumber\\
%%%%%%%%%%%%%%%%%%%%%%%%%%%
p^{S,cc,(1)}_{14}(z) &= \left(\frac{2}{3}+\frac{4 N_H}{27}+\frac{8 N_L}{27}+\frac{8 N_V}{27}\right) \logOne-\frac{16}{27} \logTwo+\left(-\frac{40}{27}+\frac{8 N_L}{9}+\frac{16 N_V}{9}+\frac{8 \pi ^2}{9}\right) z \nonumber \\
& +\frac{224}{81}+\frac{28 N_L}{81}+\frac{28 N_V}{81}+\frac{2 \pi }{27 \sqrt{3}}+\frac{4 \pi ^2}{9}+N_H \left(\frac{68}{81}-\frac{4 \pi }{9 \sqrt{3}}\right), \nonumber\\
%%%%%%%%%%%%%%%%%%%%%%%%%%%
p^{S,cc,(1)}_{15}(z) &= -\frac{64}{3} \logOne-\frac{1024}{9} \logTwo-\frac{27952 z}{9}-\frac{2128}{9}-\frac{64 \pi }{9 \sqrt{3}}, \nonumber\\
%%%%%%%%%%%%%%%%%%%%%%%%%%%
p^{S,cc,(1)}_{16}(z) &= \left(\frac{32}{3}+\frac{40 N_H}{27}+\frac{80 N_L}{27}+\frac{80 N_V}{27}\right) \logOne-\frac{256}{27} \logTwo \nonumber \\
& +\left(-\frac{1720}{27}+\frac{80 N_L}{9}+\frac{160 N_V}{9}+\frac{128 \pi ^2}{9}\right) z \nonumber \\
& +\frac{2344}{27}+\frac{208 N_L}{81}+\frac{208 N_V}{81}+\frac{32 \pi }{27 \sqrt{3}}+\frac{64 \pi ^2}{9}+N_H \left(\frac{608}{81}-\frac{40 \pi }{9 \sqrt{3}}\right), \nonumber\\
%%%%%%%%%%%%%%%%%%%%%%%%%%%
p^{S,cc,(1)}_{23}(z) &= 8 \logOne-\frac{16}{3} \logTwo-\frac{448 z}{3}+ \frac{116}{9}+\frac{8 \pi }{3 \sqrt{3}}, \nonumber\\
%%%%%%%%%%%%%%%%%%%%%%%%%%%
p^{S,cc,(1)}_{24}(z) &= \left(8-\frac{8 N_H}{9}-\frac{16 N_L}{9}-\frac{16 N_V}{9}\right) \logOne+\frac{32}{9} \logTwo \nonumber \\
&+\left(\frac{728}{9}-\frac{16 N_L}{3}-\frac{32 N_V}{3}-\frac{16 \pi ^2}{3}\right) z \nonumber \\
& + \frac{632}{27}-\frac{56 N_L}{27}-\frac{56 N_V}{27}-\frac{4 \pi }{9 \sqrt{3}}-\frac{8 \pi ^2}{3}+N_H \left(-\frac{136}{27}+\frac{8 \pi }{3 \sqrt{3}}\right), \nonumber\\
%%%%%%%%%%%%%%%%%%%%%%%%%%%
p^{S,cc,(1)}_{25}(z) &= 128 \logOne-\frac{256}{3} \logTwo -\frac{6304 z}{3}+\frac{32}{3}+\frac{128 \pi }{3 \sqrt{3}},\nonumber\\
%%%%%%%%%%%%%%%%%%%%%%%%%%%
p^{S,cc,(1)}_{26}(z) &= \left(128-\frac{80 N_H}{9}-\frac{160 N_L}{9}-\frac{160 N_V}{9}\right) \logOne+\frac{512}{9} \logTwo \nonumber \\
& +\left(\frac{9920}{9}-\frac{160 N_L}{3}-\frac{320 N_V}{3}-\frac{256 \pi ^2}{3}\right) z \nonumber \\
&+\frac{2800}{9}-\frac{416 N_L}{27}-\frac{416 N_V}{27}-\frac{64 \pi }{9 \sqrt{3}}-\frac{128 \pi ^2}{3}+N_H \left(-\frac{1216}{27}+\frac{80 \pi }{3 \sqrt{3}}\right)\,,
\label{eq::pS_ij^(1)}
\end{align}
with $\logOne = \log(\mu_1^2/m_b^2)$ and $\logTwo = \log(\mu_2^2/m_b^2)$.
Furthermore, we introduce the symbols $N_L$, $N_V$ and $N_H$
which label closed fermion loops with mass $0$, $m_c$ and $m_b$,
respectively. In the numerical evaluation we set $N_L=3$, $N_V=1$ and
$N_H=1$.

The results for $p_{ij}^{uu}$ and $p_{ij}^{S,uu}$ are obtained from
$p_{ij}^{cc}$ and $p_{ij}^{S,cc}$ for $z=0$.
For $p_{ij}^{uc}$ and $p_{ij}^{S,uc}$ we have
\begin{align}
	p^{uc,(0)}_{ij}(z) &= \frac{p^{cc,(0)}_{ij}(z) + p^{cc,(0)}_{ij}(0)}{2}, \quad 
	p^{S,uc,(0)}_{ij}(z) = \frac{p^{S,cc,(0)}_{ij}(z) +
                             p^{S,cc,(0)}_{ij}(0)}{2}\,.
\end{align}
Since we perform an expansion up to linear order in $z$, 
the NLO coefficients $p^{uc,(1)}_{ij}(z)$ can 
be cast in the following compact form
\begin{align}
p^{uc,(1)}_{13}(z) &= p^{cc,(1)}_{13}(z/2), \nonumber \\
p^{uc,(1)}_{14}(z) &= p^{cc,(1)}_{14}(z/2) + \frac{5}{18} z N_V, \nonumber \\
p^{uc,(1)}_{15}(z) &= p^{cc,(1)}_{15}(z/2) - 384 z \log(2), \nonumber \\
p^{uc,(1)}_{16}(z) &= p^{cc,(1)}_{16}(z/2) + \frac{25}{9} z N_V + 16 z \log (2), \nonumber \\
%%%
p^{uc,(1)}_{23}(z) &= p^{cc,(1)}_{23}(z/2), \nonumber \\
p^{uc,(1)}_{24}(z) &= p^{cc,(1)}_{24}(z/2) - \frac{5}{3} z N_V, \nonumber \\
p^{uc,(1)}_{25}(z) &= p^{cc,(1)}_{25}(z/2) - 288 z \log(2), \nonumber \\
p^{uc,(1)}_{26}(z) &= p^{cc,(1)}_{26}(z/2) - \frac{50}{3} z N_V - 96 z \log (2),
\label{eq:p26}
\end{align}
as well as 
\begin{align}
p^{S,uc,(1)}_{13}(z) &= p^{S,cc,(1)}_{13}(z/2), \nonumber \\
p^{S,uc,(1)}_{14}(z) &= p^{S,cc,(1)}_{14}(z/2) + \frac{4}{9} z N_V, \nonumber \\
p^{S,uc,(1)}_{15}(z) &= p^{S,cc,(1)}_{15}(z/2) , \nonumber \\
p^{S,uc,(1)}_{16}(z) &= p^{S,cc,(1)}_{16}(z/2) + \frac{40}{9} z N_V, \nonumber \\
%%%
p^{S,uc,(1)}_{23}(z) &= p^{S,cc,(1)}_{23}(z/2), \nonumber \\
p^{S,uc,(1)}_{24}(z) &= p^{S,cc,(1)}_{24}(z/2) - \frac{8}{3} z N_V, \nonumber \\
p^{S,uc,(1)}_{25}(z) &= p^{S,cc,(1)}_{25}(z/2) , \nonumber \\
p^{S,uc,(1)}_{26}(z) &= p^{S,cc,(1)}_{26}(z/2) - \frac{80}{3} z N_V. \label{eq:ps26}
\end{align}
The expressions in Eqs.~(\ref{eq::p_ij^(1)}) to~(\ref{eq:ps26}) are exact to
order $z$, i.e.\ they receive corrections of order $z^2\log z$.  
Computer-readable expressions of the two-loop coefficients from
Eqs.~(\ref{eq::p_ij^(1)}),~(\ref{eq::pS_ij^(1)}),~(\ref{eq:p26})
and~(\ref{eq:ps26}) can be found in the ancillary file to this
paper~\cite{progdata}.  The two-loop terms proportional to $N_L$, $N_V$ and
$N_H$ have recently been computed in Ref.~\cite{Asatrian:2020zxa} and we find
complete agreement after expanding the exact expression up to linear order in
$z$ and transforming to the operator basis used in~\cite{Asatrian:2020zxa}.
We note that the NLO coefficients with $i=1,2$ and $j=8$ are only one-loop
quantities and can be extracted from Ref.~\cite{Beneke:1998sy}.

It is interesting to note that the $\sqrt{3}$ in our results originate from
the Feynman diagrams in Fig.~\ref{fig::sample}(b) where in one of the closed
loops a massive bottom quark is present.  We mention that our results passes
the checks mentioned above, the gauge parameter and the gluon mass vanish. As
an additional check we have re-done the calculation employing dimensional
regularisation of the IR divergences, which requires to do the LO matching at
order $\epsilon$, and found the same results.

The results in Eqs.~(\ref{eq::p_ij^(1)}) to~(\ref{eq:ps26}) contain
terms of order $z\log z$ which result from diagrams with charm
self-energies and mass counterterms. The large coefficients of these
terms, proportional to the LO term $\gamma_m^{(0)}=6 C_F=8$ of the mass
anomalous dimension, weakens the quality of the perturbative expansion
and is especially troublesome for the prediction of $a_{\rm fs}^q$,
from which the $z^0$ terms cancel. To eliminate these terms one
employs the one-loop relation
\begin{eqnarray}
  z &=&  \bar z \,
  \lt(1- \gamma_m^{(0)}\frac{\alpha_s(m_b)}{4\pi}
  \log {\bar{z}} \rt) +{\cal O}
  (\alpha_s^2)\,, \nonumber\\
  p_{ij}^{ab}(z) 
  &=&  p_{ij}^{ab}\lt( \bar z\rt) - \frac{\partial
    p_{ij}^{ab,(0)}(\bar z) }{\partial \bar z} \,
  \frac{\alpha_s(m_b)}{4\pi}  \, 
  \gamma_m^{(0)}  \bar z \log\bar z +{\cal O}
  (\alpha_s^2) 
\end{eqnarray}  
so that trading $z$ for $\bar z$ requires the replacement
\begin{eqnarray}  
   p_{ij}^{ab,(1)} (z) &\to&   \bar p_{ij}^{\,ab,(1)} (\bar z) \;\equiv \;
                             p_{ij}^{ab,(1)} (\bar z) -
                             \frac{\partial
                      p_{ij}^{ab,(0)}(\bar z) }{\partial \bar z} \, 
                      \gamma_m^{(0)} \bar z \log
                             \bar z, \label{eq:repl}
\end{eqnarray}  
where $\alpha_s(m_b)=\alpha_s(\mu_1) + {\cal O} (\alpha_s^2)$ has been
used, and an analogous replacement for $ p_{ij}^{S,ab,(1)} (z)$.

The benefit of using $\bar z$ instead of $z$ for the quality of the
prediction has been demonstrated in
Refs.~\cite{Beneke:2002rj,Lenz:2006hd} and we refrain from using $z$
in our numerics. This leaves two plausible renormalisation schemes:
One may either use $(m_b^{\rm OS})^2$ or $\bar m_b^2(\bar m_b)$ in the
prefactor of the square bracket of $\Gamma_{12}^{ab}$ in
\eq{eq::Gam^ab}. The latter choice requires the replacement
\begin{eqnarray}  
  \bar p_{ij}^{\,ab,(1)} (\bar{z}) &\to&  
                                   \bar{\bar p}_{ij}^{\,ab,(1)} ( {\bar{z}} ) \; \equiv \;
                                   \bar p_{ij}^{\,ab,(1)} (\bar z) +
                                   {8 C_F} p_{ij}^{ab,(0)} (\bar z),
                                   \label{eq:repl2}
\end{eqnarray}  
and an analogous change of $\bar p_{ij}^{\,S,ab,(1)} (\bar{z})$. In
Refs.~\cite{Asatrian:2017qaz,Asatrian:2020zxa} the two mentioned schemes
are referred to as ``pole'' and ``$\ov{\rm MS}$''.

Let us next investigate the numerical effects of the new contributions to
$H^{ab}(z)$ and $\widetilde{H}^{ab}_S(z)$. For the input values we use
$\alpha_s(M_Z)=0.1179$~\cite{Zyla:2020zbs} and the $\overline{\rm MS}$ quark
masses $m_c(3~\mbox{GeV})=0.993$~GeV~\cite{Chetyrkin:2017lif} and
$m_b(m_b)=4.163$~GeV~\cite{Chetyrkin:2010ic} which leads $m_c(m_b)=0.929$~GeV
and $\bar{z}\approx 0.0497$.  From $m_b(m_b)$ we obtain
$m_b^{\rm OS}=4.56$~GeV using the one-loop conversion formula. For the
computation of the $\Delta B=1$ matching coefficients we use as matching scale
to the Standard Model $\mu_0=M_W=80.403$~GeV. The scale $\mu_1$ is set to
$m_b(m_b)$.

In the following we discuss the ``cc'' contribution of the quantities $H^{ab}$
and $\widetilde{H}^{ab}_S$ in the $\overline{\rm MS}$ scheme.  We refrain from
showing explicit results for the ``uu'' and ``uc'' contributions which show a
similar pattern.  We have
\begin{eqnarray}
  H^{cc} \!\!\!
  &=& \!\!\!
      0.925_{(c)} - 0.051_{(cp)} + (0.002 N_V + 0.002 N_L)_{(p)}
      \nonumber\\\mbox{}
  &&
     \!\!\!+\frac{\alpha_s}{4\pi}\bigg[
     - 2.566_{(c)}
     - 0.696_{(c-gb)}
     + (-0.846 
     + 0.0128 N_H
     + 0.116 N_V
     + 0.105 N_L)_{(cp)}
             \bigg]\,,
     \nonumber\\
  \widetilde{H}^{cc}_S \!\!\!
  &=& \!\!\!
      1.606_{(c)} - 0.084_{(cp)} + (0.002 N_V + 0.002 N_L)_{(p)}
      \nonumber\\\mbox{}
  &&
     \!\!\!+\frac{\alpha_s}{4\pi}\bigg[
     - 0.791_{(c)}
     - 1.114_{(c-gb)}
     + (-1.363
     + 0.021 N_H
     + 0.186 N_V
     + 0.168 N_L)_{(cp)}
             \bigg]\,,
     \nonumber\\
  \label{eq::Hcc_MS}
\end{eqnarray}
where ``{\it c-gb}'' refers to the diagrams with two current-current operators
and a gluon bridge, see Fig.~\ref{fig::sample}(c). The numerical values are
specific to the operator renormalisation scheme chosen by us. The scheme
dependence cancels in combination with the NLO Wilson coefficients $C_{3-6}$
entering the numbers label with ``{\it cp}''.  From Eq.~(\ref{eq::Hcc_MS}) we
observe that at one-loop order the penguin contribution is about a factor 20
smaller than the terms proportional to $C_1$ and $C_2$, which justifies
to calculate penguin contributions to lower orders in $\alpha_s$ than those
with two copies of $C_{1,2}$.  However, at two loops the impact of the
penguin coefficients is larger. In the case of $H^{cc}$ the relative
factor is less than three and in the case of $\widetilde{H}^{cc}_S$ the
penguin coefficient is even bigger than the current-current contribution. We
want to remark that the numerically most important penguin contribution is the
one proportional to $C_4$.

We want to remark that in all cases the fermionic contributions to the
the penguin coefficients, which are known from Ref.~\cite{Asatrian:2020zxa},
are significantly smaller than the non-fermionic terms computed
in this paper. Still, using $N_H=N_V=1$ and $N_L=3$ we observe a
screening of the non-fermionic coefficient of close to 50\%.

We observe that the expansion in $z$ is well-behaved. For example,
more than 90\% of the non-fermionic penguin coefficients at two-loop order
in Eq.~(\ref{eq::Hcc_MS}) are provided by the $m_c\to 0$ approximation.

We are now in the position to evaluate the shift of the new
corrections to the width difference. To illustrate the numerical effect
of the new corrections we omit both the fermionic NNLO contributions
computed in~\cite{Asatrian:2017qaz} and power corrections of order
$\Lambda_{\rm QCD}/m_b$. We furthermore concentrate on
$\Delta\Gamma_s$. In addition to the quark masses and $\alpha_s$ given
above we have the following input
parameters~\cite{Tanabashi:2018oca,Dowdall:2019bea,Bazavov:2017lyh}
\begin{eqnarray}
  M_{B_s} &=& 5366.88\, \mbox{MeV}\,,\nonumber\\ 
  B_{B_s} &=& 0.813 \pm 0.034\,,\nonumber\\      
  \widetilde{B}_{S,B_s}^\prime &=& 1.31 \pm 0.09\,,\nonumber\\
  f_{B_s} &=& (0.2307 \pm  0.0013)\, \mbox{GeV}\,,\nonumber\\ 
  \frac{\lambda_u^s}{\lambda_t^s} &=&
                                      - (0.00865 \pm 0.00042)
                                      + (0.01832\pm 0.00039) i\,.
\end{eqnarray}

\begin{table}[t]
  \begin{center}
    \begin{tabular}{|c|c|c|}
      \hline 
      Correlator & Perturbative order & $z$-dependence \\ 
      \hline 
      $O_{1,2} \times O_{1,2}$ \cite{Beneke:1998sy}  & 1 loop & exact \\
      $O_{1,2} \times O_{1,2}$ \cite{Beneke:1998sy}  & 2 loops & exact \\
      $O_{1,2} \times O_{8}$ \cite{Beneke:1998sy} & 1 loop & exact \\
      $O_{1,2} \times O_{3-6}$ \cite{Beneke:1998sy}  & 1 loop & exact \\
      $O_{1,2} \times O_{3-6}$              & 2 loops & $\mathcal{O}(z)$ \\
      $O_{3-6} \times O_{3-6}$ \cite{Beneke:1996gn}  & 1 loop & exact\\
      \hline 
    \end{tabular} 
    \caption{List of ingredients relevant for $\Delta\Gamma_s$.
      The two-loop result for the $O_{1,2} \times O_{3-6}$
      contribution is new.}
    \label{tab:nlo-ingr}
  \end{center}
\end{table}

Let us first consider the quantity $\Delta\Gamma_s$. The contributions
entering our prediction are explicitly listed in table \ref{tab:nlo-ingr}
Including all known NLO corrections we obtain
\begin{eqnarray}
  \Delta\Gamma_s &=& 0.105~\mbox{ps}^{-1} 
                     + \ldots
                     \quad ({\rm pole})\,,\nonumber\\
  \Delta\Gamma_s &=& 0.110~\mbox{ps}^{-1} 
                     + \ldots
                     \quad (\overline{\rm MS})\,,
\end{eqnarray}
where the ellipses indicate terms of order $\Lambda_{\rm QCD}/m_b$.
In case the new corrections computed in this paper are excluded we
have
\begin{eqnarray}
  \Delta\Gamma_s &=& 0.108~\mbox{ps}^{-1} + \ldots \quad ({\rm pole})\,,\nonumber\\
  \Delta\Gamma_s &=& 0.113~\mbox{ps}^{-1} + \ldots \quad (\overline{\rm MS})\,.
\end{eqnarray}
Thus the calculated corrections increase $\dg_s$ by $0.003\,\mbox{ps}^{-1}$,
which is almost as large as today's experimental error in \eq{eq:exp}. The
size of the correction is also in the ballpark of the hadronic uncertainty, if
$\dg_s$ is predicted from $\dg_s/\dm_s$, since hadronic uncertainties largely
cancel from this ratio \cite{Lenz:2006hd,Asatrian:2020zxa}.
  
Next, we discuss the relative shift of $\Delta\Gamma_s$ due to the
contribution of the penguin contribution in more detail. At one-loop
order we obtain
\begin{eqnarray}
  \frac{\Delta\Gamma_s^{p,12\times36,\alpha_s^0}}{\Delta\Gamma_s}
  &=& 7.6 \% \quad ({\rm pole})\,,\nonumber\\
  \frac{\Delta\Gamma_s^{p,12\times36,\alpha_s^0}}{\Delta\Gamma_s}
  &=& 6.1 \% \quad (\overline{\rm MS})\,,
      \label{eq::ratio_0}
\end{eqnarray}
where the quantity in denominator includes all current-current and
current-penguin corrections up to order $\alpha_s^1$. The
penguin-penguin contributions are included up to order $\alpha_s^0$
(one-loop order).
The numerator in Eq.~(\ref{eq::ratio_0}) only contains the
LO current-penguin contributions (indicated by the superscript
``$12\times36$''). 

At two-loop order we have
\begin{eqnarray}
  \frac{\Delta\Gamma_s^{p,12\times36,\alpha_s}}{\Delta\Gamma_s}
  &=& 0.3 \% \quad ({\rm pole})\,,\nonumber\\
  \frac{\Delta\Gamma_s^{p,12\times36,\alpha_s}}{\Delta\Gamma_s}
  &=& 1.4 \% \quad (\overline{\rm MS})\,,\nonumber\\
\end{eqnarray}
where the numerator contains the new corrections computed in this paper
together with the corresponding fermion contributions
from~\cite{Asatrian:2020zxa}.
Note that the non-$N_f$ penguin contribution overcompensates the
$N_f$ terms~\cite{Asatrian:2020zxa}. In the pole scheme 
this leads to tiny corrections below the percent level.
In the $\overline{\rm MS}$ scheme the non-$N_f$ contribution is about
a factor three bigger than the $N_f$ terms which leads to a
relative correction of $-1.4 \%$.

%- }}}
%- {{{ Conclusions:

\section{\label{sec::concl}Conclusions}

In this paper, for the first time, the $\Delta B=1$ operator basis from
Ref.~\cite{Chetyrkin:1997gb} has been used for the computation of NLO
corrections to the decay matrix element $\Gamma_{12}^q$, governing the width
difference between the eigenstates of the \bbq\ mass matrix and the CP
asymmetry in semileptonic $B_q$ decays.
After reproducing known results
\cite{Beneke:1998sy,Beneke:2003az,Ciuchini:2003ww,Asatrian:2017qaz,Asatrian:2020zxa}
we have obtained novel two-loop contribution to $\Gamma_{12}^q$, namely all
contributions involving one current-current operator and one four-quark
penguin operator.  We have computed these two-loop
corrections
in an expansion in $m_c/m_b$ including quadratic
terms. Computer-readable expressions of our results can be downloaded from~\cite{progdata}.

The calculated NLO effects dominate over the previously known partial results
which contain only fermion loop contributions. While the NLO penguin
contributions are numerically less relevant than those with two large
current-current coefficients $C_{1,2}$, they are needed for the theory
prediction to match the experimental precision of
$\dg_s$ in \eq{eq:exp}. To fully keep up with experiment one further needs the
contributions involving $Q_8$ at the two-loop level and a full NNLO (three-loop)
calculation of the contributions with two current-current operators. For the
NNLO calculation it is instrumental to use the CMM operator
basis~\cite{Chetyrkin:1997gb} as we did in this paper. 

%- }}}

%- {{{ Acknowledgements:

\section*{Acknowledgements}  

We thank Artyom Hovhannisyan for providing to us intermediate results of
Ref.~\cite{Asatrian:2017qaz}.  This research was supported by the Deutsche
Forschungsgemeinschaft (DFG, German Research Foundation) under grant 396021762
--- TRR 257 ``Particle Physics Phenomenology after the Higgs Discovery''.

%- }}}

\begin{appendix}

%- {{{ App. A: Projector:

\section{\label{app::proj} Projector methodology}

In this appendix we briefly describe the approach based on the construction of
projectors for the various tensor structures.  In general, the scattering
amplitude of the process $ b + \bar{s} \rightarrow \bar{b} + s$ can be
parametrized as
\begin{equation}
    \begin{aligned}
        \mathcal{M} &= \sum_{n,m} \ A^{(n,m)} \ \Gamma_{i_1i_2i_3i_4}^{(n)} \
        \Sigma_{c_1c_2c_3c_4}^{(m)} \ \bar{s}_{i_1}^{c_1} b_{i_2}^{c_2} \
        \bar{s}_{i_3}^{c_3} b_{i_4}^{c_4}\,, 
    \end{aligned}
    \label{eq:basis-decomposition}
\end{equation}
where $c_n$ describe the colour and $i_n$ the spinor indices. Note that
the number of colour structures $\Sigma^{(m)}$ is finite.
On the other hand, the basis of the Lorentz
structure is \textit{a priori} not finite. For a massless $s$
quark the Lorentz structure can be expressed
as\footnote{For non-SM interactions or with massive $s$-quarks the
  generalization to arbitrary chiralities is straightforward.}
\begin{equation}
  \begin{aligned}
    \Gamma^{(n)}_{i_1i_2i_3i_4} = \left(P_R B^{(n)} \right)_{i_1i_2} \
    \left(P_R B^{(n)} \right)_{i_3i_4} \equiv \left(P_R B^{(n)} \right)
    \otimes \left(P_R B^{(n)} \right) \,,
  \end{aligned}
  \label{eq:lorentz-decomposition}
\end{equation}
where $P_R=(1+\gamma_5)/2$ and the basis vectors $B^{(n)}$ are given by
\begin{align}
  B^{(0)} = \mathbf{1}\,, \quad B^{(1)} = \gamma_{\mu_1} \,, \quad
  B^{(2)} = \gamma_{\mu_1} \gamma_{\mu_2} \,, \quad B^{(3)} =
  \gamma_{\mu_1} \gamma_{\mu_2} \gamma_{\mu_3} \,, \dots\, .
  \label{eq:basis-vectors}
\end{align}
In four space-time dimensions it is possible to avoid chains with 
more than four Dirac matrices, which is not possible in $d=4-2\epsilon$
dimensions. However, in a fixed order in the perturbative expansion only a
finite number of basis vectors $B^{(n)}$ appear.

In general, the coefficients $A^{(n,m)}$ include dimensionally regularized scalar Feynman
integrals. To extract $A^{(n,m)}$ in \eq{eq:basis-decomposition}, one
can apply tensor reduction to get scalar integrand expressions. Alternatively,
one can make use of the composition of \eq{eq:basis-decomposition}. Hence, we
define projection operators for Lorentz ($\mathbf{P}^{(n)}$) and 
colour space ($\mathbf{C}^{(m)}$), acting as
\begin{align}
    A^{(n,m)} = \mathbf{C}^{(m)} {\mbox{Tr}_d\left[\mathbf{P}^{(n)} \mathcal{M}\right]\,,}
  \label{eq::Anm}
\end{align}
where $\mathbf{C}^{(m)}$ commutes with the operations applied in
Lorentz space.  $\mathbf{P}^{(n)}$ is constructed from a linear combination of
the structures introduced in Eq.~(\ref{eq:basis-vectors}). It is understood
that the traces are evaluated in $d$ dimensions. Note that in our case
no traces including $\gamma_5$ appear since \eq{eq:lorentz-decomposition}
explicitly contains a projector $P_R$. Thus, Eq.~(\ref{eq::Anm}) takes the form
\begin{equation}
    \begin{aligned}
      A^{(n,m)} = \ \mathbf{C}^{(m)} \sum_{i} p^{(n,i)} \ \mbox{Tr}_d\bigg[ \left(B^{(i)} \otimes B^{(i)}
      \mathcal{M}\right) \bigg]\,, \nonumber
    \end{aligned}
\end{equation}
Using the explicit structure of $\mathcal{M}$ we can express the projector
coefficients $p^{(i,j)}$ as the inverse of the Gram matrix, constructed from
the tensor basis of \eq{eq:lorentz-decomposition}
\begin{eqnarray}
  \left(p^{-1}\right)^{(i,j)} &=& \mbox{Tr}_d\left[B^{(i)} B^{(j)} \otimes
                                B^{(i)} B^{(j)} \right]\,.
\end{eqnarray}
Note that on the right-hand side one has a product of two traces.

A caveat of this approach is that the complexity of the matrix $p$ grows
considerably with the number of $\gamma$ matrices in the basis elements
$B^{(n)}$. For our NLO calculation, we have to consider terms up to $n=9$
which leads to products of two $d$-dimensional traces where each one contains
up to 18~$\gamma$-matrices. This non-trivial computational task was done using
\texttt{FORM}~\cite{Kuipers:2012rf}, where the special hints described in the
manual have been used. To avoid unnecessary recomputations, we evaluate each
occurring trace product separately and include the result in a lookup table.

%- }}}

\end{appendix}

%- {{{ Bibl.:

%- }}}

\end{document}